V. I. Skalozubov[1], V. N. Vashchenko[2], S. S. Jarovoj[2], V. Yu. Kochnyeva[1]

[1] *Institute for Safety Problems of Nuclear Power Plants, National Academy of Sciences of Ukraine, Kyiv*
[2] *State Ecological Academy for Postgraduate Education and Management, Kyiv*

# SUBSTANTIATION OF THERMODYNAMIC CRITERIA OF EXPLOSION SAFETY IN PROCESS OF SEVERE ACCIDENTS IN PRESSURE VESSEL REACTORS

The paper represents original development of thermodynamic criteria of occurrence conditions of steam-gas explosions in the process of severe accidents. The received results can be used for modelling of processes of severe accidents in pressure vessel reactors.

In the process of a severe accident (SA) at WWER explosions of hydrogenous steam-air mixtures (HSAMs) and steam (power) explosions are the most probable concerning conditions of occurrence and dangerous concerning consequences. The analysis shows necessity of the substantiation (additional to thermochemical) criteria of explosion safety of the steam-gas mixtures considering specificity of thermohydrodynamic conditions and mechanisms of formation of explosive situations at different stages of SA development.

The dominant sources of formation of gaseous hydrogen in SA process are zirconium-steam reaction and consequences of steam (power) explosions.

The rate of zirconium-steam reaction (generation of gaseous hydrogen mass per time unit) is generally defined by temperature (specific enthalpy) fuel-containing materials (FCM) and becomes essential at the temperatures exceeding damage conditions of zirconium-containing fuel element claddings (above 1200 ºC). Therefore, assurance of conditions of stable decrease in FCM temperature is the dominant factor of decrease in mass and concentration of generated gaseous hydrogen and, consequently, one of conservative thermodynamic criteria of explosion safety of HSAMs:

$$\frac{dT_{FCC}}{dt}, \frac{di_{FCC}}{dt} < 0 \qquad (1)$$

Where $T_{FCM}$, $i_{FCM}$ are temperature and specific enthalpy of FCM; $t$ is time.

Conservatism of criterion is defined by that in case of default of a condition (1) there will be explosion of HSAM; i.e. it is considered in these conditions local concentrations of hydrogen and oxygen will be enough for occurrence of processes of a deflagration/detonation regardless of concentration of water steam, nitrogen, inert gases and others recombiners.

In addition, the important factors defining conditions of explosion safety of HSAMs are the mass and concentration of the water steam, which is per se decatalyzer of explosion of HSAM. Therefore assurance of conditions of stable keeping of mass of water steam $M_S$, and accordingly concentration of steam-gas mixture is the second conservative criterion of explosion safety of HSAM:

$$\frac{dM_S}{dt} > 0. \qquad (2)$$

Rates of temperature change of FCMs and mass of water steam are defined by intensity of processes of heat-mass exchange and heat-mass transfer between FCMs, the coolant and a steam-gas mixture, and conditions of the organisation of cooling in process of SA development also.

The basic criterion of occurrence of steam (power) explosion is if rate of pressure increasing in steam-gas volume $dP/dt$ exceeds corresponding critical values of a power detonation of steam-gas mixture:

$$\frac{dP}{dt} \geq \left(\frac{dP}{dt}\right)_{cr} = P'_{cr}. \qquad (3)$$

Values $P'_{cr}$ are defined by local physical and chemical properties of steam-gas medium in the conditions of a possible detonation. For conditions of SA at WWER, experimentally confirmed conservative estimation $P'_{кр}$ is $10^3$ MPa/s [1].

Rate of pressure increasing in steam-gas volume is defined by intensity of processes of heat-mass exchange and heat-mass transfer in the conditions of a possible detonation (in this case – in the conditions of multiphase thermohydrodynamic interaction of nuclear fuel, structures, coolant, sources of cooling and steam-gas medium during development of SA in the equipment/systems of pressure vessel reactors).

For an in-vessel or out-vessel stage of SA, the thermohydrodynamic substantiation of conditions of occurrence of steam-gas explosions is fulfilled using model of the reduced volume containing concentrated steam-gas volume $V_{SG}$ and coolant pool with volume $V_L$ and fuel melt mass $M_M$ (see drawing). Feeding of cooling water to the reduced volume from safety systems is modelled by a source with the total flow $G_{LT}$ and specific enthalpy (temperature) $i_T(T_T)$. Generally, removal of steam-gas medium through the organized and unorganized leakages of the reduced volume are modelled by total flow $G_{leak}$.

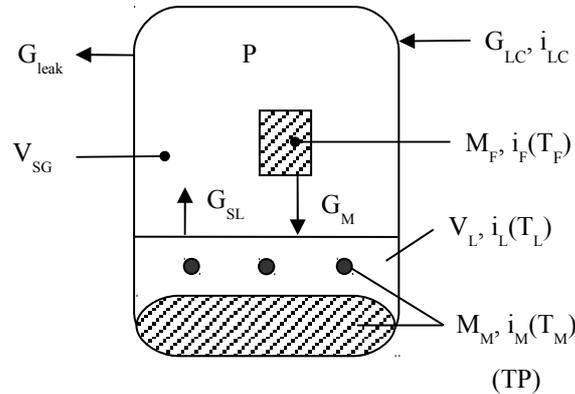

(TP)
Thermohydrodynamic model of conditions of steam-gas explosions in process of SA.

The main sources of hydrogen and heat (including because of zirconium-steam reaction) are modelled by concentrated FCMs in steam-gas volume (with the reduced mass $M_F$ and specific enthalpy $i_F(T_F)$) and melt of FCMs in coolant pool $(M_M, i_M)$. It is conservatively considered that concentrated FCMs have the maximum temperatures of the most heat-stressed elements and corresponding thermophysical properties.

Taking into account the accepted assumptions, the thermohydrodynamic model of conditions of occurrence of steam-gas explosions is:

$$\frac{dM_S}{dt} = G_{SL} - G_{leak} - G_{con}, \qquad (4)$$

$$\frac{dM_F}{dt} = -G_M, \qquad (5)$$

$$\frac{dM_{LC}}{dt} = G_{LC} + G_{con}, \qquad (6)$$

$$\frac{dM_S i_S}{dt} = G_{SL} i_S + q_{SF} \Pi_{SF} - q_{con} \Pi_{con}, \qquad (7)$$

$$\frac{dM_F i_F}{dt} = -q_{SF} \Pi_{SF} + Q_V(i_F) - q_{LCF} \Pi_{LCF} - G_M i_F, \qquad (8)$$

$$\frac{dM_{LC}i_{LC}}{dt} = q_{con}\Pi_{con} + q_{LCF}\Pi_{LCF},\qquad(9)$$

$$\frac{dM_L}{dt} = -G_{SL} + G_{LC} + G_{con},\qquad(10)$$

$$\frac{dM_M}{dt} = G_M,\qquad(11)$$

$$\frac{dM_L i_L}{dt} = -G_{SL}i + q_{LM}\Pi_{LM} + (G_{LC} + G_{con})i_{LC},\qquad(12)$$

$$\frac{dM_M i_M}{dt} = G_M i_F - q_{LM}\Pi_{LM} + Q_{VM}(i_M),\qquad(13)$$

Where $G_{SL}$ – steam generation flow in pool with liquid; $G_{con}$ – mass flow of condensed steam; $G_{leak}$ – flow through volume leakage; $M_M$, $G_M$ – mass and flow of fuel melt in liquid pool; $i_L$, $i_S$, $i_F$, $i_M$ – specific enthalpy of liquid, steam, fuel structures in steam-gas volume and melt in liquid pool; $q_{LM}$, $q_{FS}$ – density of heat flows between liquid/melt and fuel/steam; $\Pi_{LM}$, $\Pi_{FS}$ – area of contact between liquid/melt and fuel elements/steam; $Q_{VF}$, $Q_{VM}$ – specific internal sources of thermal energy of fuel structures in steam-gas volume and melt; $M_F$ – mass of fuel elements in steam-gas volume; $M_{LC}$, $i_{LC}$, $G_{LC}$ – mass, specific enthalpy, and total flow of cooling source; $\alpha_L$, $\alpha_S$ – heat-transfer factor between FCM and liquid and steam; $q_{con}$ – density of heat flow during steam condensation; $\Pi_{con}$ – contact area of condensation; $q_{LCF}$, $\Pi_{LCF}$ – density of heat flow and contact area of heat transfer between FCM in steam-gas volume and cooling source.

Boundary conditions of interphase heat transfer are

$$G_{con}(i_S - i_{LC}) = q_{con}\Pi_{con},\qquad(14)$$

$$Q_{LM} = \alpha(T_M - T_L),\qquad(15)$$

$$Q_{SF} = \alpha_S(T_F - T_S),\qquad(16)$$

$$Q_{LCF} = \alpha_{LC}(T_F - T_{LC}).\qquad(17)$$

Taking into account compressibility and thermodynamic nonequilibrium of steam

$$\frac{dM_S}{dt} = \rho_S \frac{dV_{SG}}{dt} + V_{SG}\frac{dP}{dt}\left(\frac{1}{a_S^2} + \frac{\partial\rho_S}{\partial i_S}\frac{\partial i_S}{\partial P}\right),\qquad(18)$$

Where $\rho_S$, $a_S$ – steam density and sound speed in steam; $V_{SG}$ – steam-gas volume "free" of structures.

The decision of thermohydrodynamic model (4) – (18) concerning criteria of explosion hazard of steam-gas mixture (1) – (3) is:

$$\frac{di_F}{dt} = \frac{Q_{VF}(i_F) - q_{SF}\Pi_{SF} - q_{LCF}\Pi_{LCF}}{M_F},\qquad(19)$$

$$\frac{di_M}{dt} = I_M(t),\qquad(20)$$

$$\frac{dM_S}{dt} = G_{SL}(t) - G_{leak}(t) - G_{con}(t),\qquad(21)$$

$$\frac{dP}{dt} = \frac{G_{SL}(1-\rho_S/\rho_L) - G_{con}(1-\rho_S/\rho_L) - G_{leak} + G_{LC}\rho_S/\rho_L}{V_{SG}\left(\dfrac{1}{a_S^2} + \dfrac{\partial\rho_S}{\partial i_S}\dfrac{\partial i_S}{\partial P}\right)},\qquad(22)$$

Where

$$G_{con}(T) = \frac{q_{con}\Pi_{con}}{i_S - i_{LC}}, \qquad (23)$$

$$G_{SL}(T) = \frac{C_L M_L\left(\dfrac{q_{LM}}{\alpha_L} - \dfrac{I_M}{C_M}\right) + q_{LM}\Pi_{LM} - G_{con}(i_L - i_{LC}) - G_{LC}(i_L - i_{LC})}{i_S - i_L}, \qquad (24)$$

$$I_M = \frac{G_M(i_F - i_M) + Q_{VM}(i_M) - q_{LM}\Pi_{LM}}{M_M}, \qquad (25)$$

$$M_F = M_{FC}(t=0) - \int_0^t G_M \, dt \qquad (26)$$

$$M_L = M_{LC}(t=0) + \int_0^t (G_{con} + G_{LC} - G_{SL}) \, dt, \qquad (27)$$

$$M_{LC} = \int_0^t (G_{con} + G_{LC}) \, dt, \qquad (28)$$

$$M_M = M_{MC}(t=0) + \int_0^t G_M \, dt, \qquad (29)$$

$$i_F = i_{FC}(t=0) + \int_0^t \frac{Q_{VF} - q_{SF}\Pi_{SF} - q_{LCF}\Pi_{LCF}}{M_F} \, dt, \qquad (30)$$

$$i_M = i_{MC}(t=0) + \int_0^t I_M \, dt, \qquad (31)$$

$$i_{LC} = i_{LC}(t=0) + \int_0^t \frac{(q_{con}\Pi_{con} + q_{LCF}\Pi_{LCF}) - i_{LC}(G_{con} + G_{LC})}{M_{LC}} \, dt, \qquad (32)$$

$$i_L = i_L(t=0) + \int_0^t \frac{q_{LM}\Pi_{LM} - G_{con}(i_L - i_{LC}) - G_{SL}(i_S - i_L) - G_{LC}(i_S - i_L)}{M_L} \, dt, \qquad (33)$$

Where $C_L$, $C_M$ – specific thermal capacity of liquid and melt.

Thus, conservative thermodynamic criteria of explosion safety for in-vessel or out-vessel stages of SA are

$$G_{SL}(t) \geq G_{leak}(t) + G_{con}(t), \qquad (34)$$

$$Q_{VF}(i_F) < q_{SF}\Pi_{SF} + q_{LCF}\Pi_{LCF}, \qquad (35)$$

$$Q_{VM}(i_M) < q_{LM}\Pi_{LM} - G_M(i_F - i_M), \qquad (36)$$

$$\frac{G_{SL}(1-\rho_S/\rho_L) - G_{con}(1-\rho_S/\rho_L) - G_{leak} + G_{LC}\rho_S/\rho_L}{V_{SG}\left(\dfrac{1}{a_S^2} + \dfrac{\partial \rho_S}{\partial i_S}\dfrac{\partial i_S}{\partial P}\right)P'_{cr}} < 1. \qquad (37)$$

The estimation of fulfilment of conditions of steam-gas explosions in the reactor vessel or in containment is possible based on deterministic calculated modelling of accident processes using the dependences of interphase heat-mass exchange and transfer proved for WWER conditions.

The qualitative analysis of the received criteria allows drawing following conclusions.

1. Influential parameters of conditions of steam-gas explosions are sources of exothermic chemical reactions $Q_{VF}$, $Q_{VM}$, processes of heat-mass exchange and transfer between FCM and the coolant $q_{LM}\Pi_{LM}$, $q_{LCF}\Pi_{LCF}$, conditions of organized and unorganized leakages $G_{leak}$, and intensity of condensation processes $G_{con}$ that can change at different stages of SA. Therefore, validity of conditions of steam explosion is directly connected with validity and applicability of calculated dependences to estimate conditions of interphase interaction during SA at WWER. It is needed to do verification and validation both the deterministic codes modelling behaviour of accident processes in whole, and calculated dependences of interphase interaction.

2. The basic control parameters (adjusting them we can affect feasibility of criteria of explosion safety in process of SA) are the flow of a cooling liquid from safety systems $G_{LC}$ and the flow of the organised removal of steam-gas medium $G_{leak}$.

The analysis of the received criteria defines ambiguity of influence of control parameters on explosion safety:

The increase in the flow of the organised removal of steam-gas medium, on the one hand, promotes to decrease in concentration of hydrogen in the reactor vessel or containment (increase of hydrogen explosion safety) and to decrease in rate of pressure increasing (increase of power explosion safety), and on the other hand, reduces mass (concentration) of water steam which is decatalyzer of hydrogen explosion (decrease of hydrogen explosion safety);

The increase in the flow of a cooling liquid, on the one hand, intensifies processes of interphase heat-mass exchange (including condensation) and promotes to decrease in FCM temperature (increase of explosion safety), and on the other hand, the fast intensification of FCM cooling can lead to cracking or fragmentation of the embrittled and oxidised surfaces under the large thermal stresses, to failure of steady film boiling and formation of new "bared" high-temperature surfaces of FCMs that promotes to increase zircaloy oxidation and melting and the subsequent relocation, to repeated fast increasing of temperature, pressure and hydrogen formation (decrease of explosion safety)[1].

In particular, the last described effect is one of actual issues of the known problem of explosion safety on uncertainty of reasonability of "repeated flooding" of damaged FCMs [1, 3]: until now there is no unique definiteness concerning necessity of "repeated reflooding" (cooling of damaged FCMs) to assure explosion safety.

Latest events at Fukushima-Daiichi have sharpen an actuality of uncertainty of effects of control parameters of explosion safety: all actions of the staff have been directed to increase the flow of cooling medium to the damaged fuel and preventive discharges of steam-gas medium; at the same time it was not possible to avoid processes of burning and a detonation that have led to destruction of the safety protective barriers and to catastrophic environmental discharges of radioactive products.

REFERENCES

1. *Skalozubov V. I., Kljuchnikov A. A, Kolyhanov V. N.* Management Basics for Beyond Design Basis Loss-of-Coolant Accident at NPPs with WWER. – Chernobyl: ISP NPP of NASU, 2010. – 400 p.
2. *Ragheb M.* Fukushima Earthquake and Tsunami Station Blackout Accident. – University of Illinois at Urbana-Champaign: NetFiles, 2011.
3. *Kabanov L. P, Kozlova N. A., Suslov A. I.* Technical Substantiation for Severe Accident Management at NPPs with WWER-1000. – SEC NRS - NRC "Kurchatov Institute", 2006.

---

[1] The known experimental researches [2] have shown that crushing of FCM fragments (for example, under the large flow of a cooling liquid) leads to decrease of heat transfer intensity almost four times.